\documentclass[12pt]{article}

\usepackage{amsmath,amssymb,cite,booktabs,multirow}

\usepackage{graphicx,color}

\newcommand{\f}[2]{\frac{#1}{#2}}
\newcommand{\hf}[2]{\mbox{\large $\frac{#1}{#2}$}}
\newcommand{\ko}[1]{\left( #1 \right)}
\newcommand{\kko}[1]{\left[ #1 \right]}

\DeclareMathOperator{\tr}{tr}

\def\pa{\partial}
\def\eq{\equiv}

\def\al{\alpha}
\def\ga{\gamma}

\def\no{\nonumber}
\def\lam{\lambda}
\def\Th{\Theta}
\def\sig{\sigma}
\def\ep{{\epsilon}}
\def\cN{{\mathcal N}}
\def\cB{{\mathcal B}}

\def\cZ{{\mathcal Z}}

\def\cD{{\mathcal D}}
\def\cW{{\mathcal W}}

\def\cO{{\mathcal{O}}}
\def\cX{{\mathcal{X}}}
\def\cY{{\mathcal{Y}}}
\def\cS{{\mathcal{S}}}

\numberwithin{equation}{section}
\begin{document}

\quad 
\vspace{-2.0cm}

\begin{flushright}
{\bf November 2009}\\
\end{flushright}

\vspace*{0.5cm}

\begin{center}
\Large\bf 
Giant Spinons
\end{center}

\vspace*{0.7cm}

\centerline{
Keisuke Okamura
}
\begin{center}
\emph{Ministry of Education, Culture, Sports, 
Science and Technology \---Japan,\\
3-2-2 Kasumigaseki, Chiyoda-ku,
Tokyo 100-8959, Japan.} \\
\vspace*{0.5cm}
\end{center}

\vspace*{1.5cm}

\centerline{\bf Abstract} 

\vspace*{0.5cm}

We study the spectrum around the ``antiferromagnetic'' states of the planar ${\rm AdS}_{5}/{\rm CFT}_{4}$ duality.
In contrast to the familiar large-spin limit $J\to \infty$ where each magnon momentum scales as $p\sim 1/J\ll 1$\,, we consider a novel ``large-winding'' limit in which the total momentum becomes infinitely large, $\sum_{j}p_{j}\to \infty$\,.
Upon taking the limit we identify ``spinon'' excitations of both gauge and string theories.
In particular, a (classical) string spinon turns out to be an infinite set of spiky strings, which are closely related to well-known infinite-spin strings\,: giant magnons.
Furthermore, we show that the curious agreement of scattering phase-shifts of two spikes and that of two giant magnons 
can be accounted for by regarding the spinon scattering as factorised scatterings of infinitely many magnons.

\vspace*{1.0cm}

\vfill

\thispagestyle{empty}
\setcounter{page}{0}

\newpage
\section{Introduction}

The AdS/CFT conjecture \cite{AdSCFT} states that the type IIB string theory on $AdS_{5}\times S^{5}$ is a dual description of the four-dimensional, $\mathcal N=4$ super Yang-Mills (SYM) theory.
A key prediction about the AdS/CFT is that the energy spectrum $\{E_n\}$ of string states with excitation levels $n$ and the spectrum of the conformal dimensions $\{\Delta_n\}$ of $\cN=4$ SYM operators match\,:
\begin{equation}
\{ E_n \} = \{ \Delta_n \}\,.
\end{equation}
In the planar limit, these quantities are supposed to be connected by some function of the 't Hooft coupling $\lambda$\,.
However, the strong/weak nature of the duality usually prevents us from a direct comparison of the spectra except for trivial BPS cases.

Nevertheless, we have been able to pursue the AdS/CFT spectrum beyond BPS in recent years with the help of numbers of significant developments in the course.
Among them probably the most striking breakthrough was the discovery of integrable structure of both theories \cite{Minahan:2002ve, Bena:2003wd, Beisert:2003tq, Beisert:2003yb, Kazakov:2004qf, Beisert:2004hm, Staudacher:2004tk, Arutyunov:2004vx, Beisert:2005fw,Beisert:2005tm,Janik:2006dc,Eden:2006rx,Arutyunov:2006iu,Beisert:2006ib,Beisert:2006ez,Arutyunov:2006yd,Dorey:2007xn}.
In the light of the AdS/CFT duality, the gauge and string integrable models must be just two ways of describing the same underlying integrable system, and it is believed that the unified integrability can be characterised by a set of Bethe ansatz equations which is valid for all values of $\lam$\,.
With the set of equations, one can compute the spectrum of the system exactly in order to obtain the interpolation energy function $E(\lam)=\Delta (\lam)$\,, although only valid in the asymptotic region ({\em i.e.}, for very long states/operators).

Below we first review the planar AdS/CFT spectrum on a ``how-far-from-BPS'' perspective, to see how far we have come on this issue.
Then we explain in which direction we can proceed, and from which direction we are going to approach, in order to achieve the programme of ``solving'' the planar ${\rm AdS}_{5}/{\rm CFT}_{4}$\,.

\subsubsection*{From Ground, or Large-Spins\,....}
At the ``bottom'' of the spectrum of $\mathcal N=4$ SYM, there exists a class of operators which are made up only of a single flavour of complex scalar field $\cZ$\,,
\begin{equation}\label{F state}
\mathcal O_{\rm F}= \tr\left( \mathcal Z^{L} \right)\,.
\end{equation} 
This is a BPS operator whose conformal dimension $\Delta$ is protected from quantum corrections\,, {\em{i.e.}}, the anomalous dimension $\Delta-L$ is zero at all values of the coupling $\lam$\,.

The dual state on the string theory side is a point-particle (collapsed closed string) circulating on one of the great circles of $S^5$ with angular momentum (or ``spin'') $L$\,.
The energy $E$ of this BPS string is equal to the large-spin $L$\,, which corresponds to the large R-charge of the SYM operator (\ref{F state}).

\paragraph{}
The study of the AdS/CFT spectrum beyond BPS was initiated by Berenstein, Maldacena and Nastase (BMN) \cite{Berenstein:2002jq}, who proposed the pp-wave/SYM correspondence.
The so-called BMN operators are obtained by replacing a small number of background $\cZ$ fields in (\ref{F state}) with other $\cN=4$ fields $\cX_{j}\in \{ \cW,\, \cY,\, F_{\mu\nu},\, \cD_{\mu},\, \Psi_{\al}^{A} \}$\,, where each field in the curly bracket denotes the second and third complex scalars, field strength, covariant derivatives and gluinos, respectively.
Schematically, the BMN operators are represented as
\begin{equation}\label{magnons}
\mathcal O_{\rm mag}\sim \tr\left( \mathcal X_{1}\dots\mathcal X_{M}\mathcal Z^{L-M} \right)+\mbox{permutations}\,,
\end{equation} 
where $M$ is the total number of impurities, which must be sufficiently small compared to the ``length'' (total number of fields in the trace) $L$ of the operators.
The impurity fields $\cX_{j}$ are also called ``magnons'' by analogy with corresponding excited states in spin-chain theories in condensed matter physics, when one regards the BPS state (\ref{F state}) as ``ferromagnetic vacuum''.\footnote{This may be a somewhat misleading naming since, from recent progress in the study of the structure of the AdS/CFT, it is indicated that the BPS states (\ref{F state}) have a hidden ``nesting'' structure.
More precisely, there seems to be hidden levels of nesting in the proposed Bethe ansatz equations of \cite{Beisert:2005fw} which becomes the origin of the nontrivial dressing phase \cite{Rej:2007vm} (see also \cite{Janik:2008hs}).
In this sense, it may be more appropriate to call the BPS state (\ref{F state}) the ``antiferromagnetic'' vacuum rather than ``ferromagnetic''.
However, for convenience, we will use these terminologies in the conventional way throughout this paper.}

Let $p_j$ be the momentum of each magnon labeled by $j$\,, then the large-spin limit of BMN is characterised as, after taking the planar limit,
\begin{equation}\label{BMN limit}
p_j\sim \frac{1}{L}\to 0\,,\qquad 
\lam\to \infty\,, \qquad
L\sim J\to \infty\,,\qquad 
\frac{\lam}{J^2}\,:\,\mbox{fixed}\ll 1\,,
\end{equation}
where $J$ is the R-charge associated with the $\cZ$ field.
In the large-spin sector, $L$ and $J$ are of the same order.
The spectrum of SYM conformal dimensions at the first few orders in the BMN coupling $\lam/J^{2}$ was shown to agree with the corresponding string energies \cite{Berenstein:2002jq, Gubser:2002tv}.

The spectrum of AdS/CFT beyond BMN (called ``far-from-BPS'' sector) was further pursued by Frolov and Tseytlin \cite{FT}, and many applications followed.
The corresponding description in spin-chain language is a classical spin-wave.

These near- and far-from-BPS sectors were for some time expected to provide an overlapping perturbative regime where one can perturbatively access from both sides of the correspondence, overcoming the strong/weak difficulty of the duality.
However, it was later proved not the case\,;
on the basis of the proposed asymptotic AdS/CFT Bethe ansatz equations, the large-spin limit (\ref{BMN limit}) is ill-defined due to the existence of a nontrivial phase factor ($\sig^{2}$ in (\ref{u4})) in the S-matrix of AdS/CFT, which breaks the BMN scaling explicitly.

\paragraph{}
A giant magnon found by Hofman and Maldacena (HM) \cite{Hofman:2006xt} is a string soliton living in a particular infinite-spin limit, and it played an important role in testing the structure of the very S-matrix of AdS/CFT.
In the HM limit, the spin $J$ and the energy $E$ of the string goes to infinity while their difference $E-J$ and the coupling $\lambda$ held fixed.
In contrast to the BMN limit (\ref{BMN limit}) in which $p\sqrt{\lam}$ is kept fixed, in the HM limit the worldsheet (or the magnon) momentum $p$ is a conserved quantity\,:
\begin{equation}\label{HM limit}
p\, :~\mbox{fixed}\,,\qquad 
\lam\, :~\mbox{fixed}\,,\qquad 
E\,,~J\to \infty\,,\qquad 
E-J\, :~\mbox{fixed}\,.\qquad 
\end{equation}
In this limit, both the gauge theory spin-chain and the dual string effectively become infinitely long, and the spectrum can then be analysed in terms of asymptotic states and their scattering.
The asymptotic spectrum of gauge theory was derived by Beisert \cite{Beisert:2005tm} by symmetry argument.
The result indeed matched with the energy-spin relation of a HM giant magnon in the strong coupling limit.
Scattering phase-shifts for two giant magnons was also derived in \cite{Hofman:2006xt}.
The results indeed reproduced the strong coupling limit of the conjectured AdS/CFT S-matrix.

\subsubsection*{...\,to Heaven, or Large-Windings}

In this paper, we explore the other end of the planar AdS${}_{5}$/CFT${}_{4}$ spectrum, that is around the ``top'' of the spectrum.
Indeed the spectrum of states/operators in AdS${}_{5}$/CFT${}_{4}$ is unbounded due to the noncompactness of the symmetry algebra $\mathfrak{psu}(2,2|4)$\,, but we can formally define the upper bound of the spectrum in the thermodynamic limit ($L\to \infty$) which should scale like $\sqrt{\lam}\,L$\,, albeit the ``length'' here is not conserved due to the nature of the dynamic spin-chain \cite{Beisert:2005tm} and thus does not have a definite value.\footnote{On this point, we will later see that the energy of the proposed antiferromagnetic state living at the ``top'' is given by $\ep_{\rm AF}\sim\sqrt{\lam}\times M$ where the number of magnons $M$ is assumed to be the same order as the spin-chain length, see (\ref{fillings2}) and (\ref{hoop}).}
Such a sector is far less understood compared to the relatively near-BPS sector, for around the top of the spectrum one cannot take standard perturbative approaches like BMN.
That is, say on the gauge theory side, solving the mixing problem is totally hopeless.

\paragraph{}
Nevertheless, we will show that one may bypass it by taking a special infinite-winding limit on both sides of the correspondence.
The novel limit we consider is the following\,:
\begin{equation}\label{LW limit}
P=\sum_{j=1}^{M}p_j
\to \infty\,, 
	\qquad \lam\, :~\mbox{fixed}\,,
	\qquad E\to \infty\,, 
	\qquad J\,,~E-gP\, :~\mbox{fixed}\,,
\end{equation}
where $g\eq \sqrt{\lam}/{4\pi}$ is the standard rescaled coupling constant.
Here $M$ is the number of magnons, which is the same order as that of the total number of fields $L$\,.
In this limit both $M$ and $L$ are sent to infinity, while the R-charge (spin) $J$ of the state is fixed.
The total momentum $P$ of the state is essentially the ``winding number'' $m\sim P/\sqrt{\lam}$ of the corresponding string state which is sent to infinity.
As usual, the momentum $P$\,, energy $E$ and spin $J$ are the three conserved charges of our interest.
Note that, in contrast to the BMN case (\ref{BMN limit}), each momentum $p_j$ of the constituent magnons is not 
necessarily of $\cO\big(\hf{1}{L}\big)$\,.

It is worthwhile to notice that the infinite-winding limit (\ref{LW limit}) is reminiscent of the standard T-duality, interchanging the spin $J$ and the winding number $m$ (only on $S^{5}$) with the infinite-spin limit (\ref{HM limit}) of HM.
We will see that in the limit (\ref{LW limit}) both string and SYM states have infinite winding number $m$\,, along one of the great circles of $S^{5}$ on the string theory side, and along a circle in the complex spectral parameter plane on the gauge theory side.
The counterpart in string theory will turn out to be the so-called ``single-spike'' string found in \cite{Ishizeki:2007we,Mosaffa:2007ty}.
From the spin-chain point of view, these states are identified as special ``spinon'' states (excitations above the antiferromagnetic vacuum) of the limiting theories, providing the first example of dual pairs around the ``top'' of the AdS${}_{5}$/CFT${}_{4}$ spectrum.

Note that the idea itself of interpreting single-spike strings as excitations above an ``antiferromagnetic'' state (in some appropriate sense) was known in the original paper \cite{Ishizeki:2007we} where the single-spikes are found.
Also, the idea of multiply-wound strings corresponding to an antiferromagnetic state was originally proposed in \cite{Roiban:2006jt} for the case of $\mathfrak{su}(2)$ sector (see also \cite{Zarembo:2005ur}).
There the arguments/approaches are all based on the relation between string sigma model on $R\times S^{3}$ and the Hubbard model \cite{Rej:2005qt}, only not totally successful.
We point out that such a sector is insufficient to encode the whole substance as is clear from the fact that the large-winding strings including single-spikes live in a non-holomorphic sector rather than holomorphic \cite{Hayashi:2007bq},\footnote{For example, consider a portion of composite SYM operator like $\cZ\overline{\cZ}$ which clearly does not live in a holomorphic sector.
It gives rise to oscillating motion in the dual sting theory picture, since if we associate $\cZ$ to a particle rotating along a great circle of $S^{5}$ say clockwise, the other particle associated with $\overline{\cZ}$ rotates counterclockwise, thus making the string connecting these two points non-rigid and oscillating.
Single-spike strings are relatives of such oscillating string as argued in \cite{Hayashi:2007bq}.}
and so that one naturally has to enlarge the scope to the full $\mathfrak{psu}(2,2|4)$\,, taking as well into account the operator mixing effect.

\paragraph{}
This paper is organised as follows.
In Section \ref{sec:BAE}, we review some relevant aspects of the asymptotic Bethe ansatz equations for the AdS/CFT system as preliminaries.
Then in Section \ref{sec:GS} we explicitly take the infinite-winding limit (\ref{LW limit}) to identify the spinon excitations of the limiting theories.
We also discuss the implication of our picture to resolve the puzzle observed in \cite{Ishizeki:2007kh} that the scattering phase-shift of single-spikes and that of giant magnons agree.
Section \ref{sec:C&D} is devoted to conclusions and discussions.

\section{Preliminaries\label{sec:BAE}}

The programme of constructing the underlying Bethe ansatz equations for the AdS/CFT spin-chain has been intensively pursued in these years \cite{Beisert:2004hm,Arutyunov:2004vx,Staudacher:2004tk}.
The full set of all-order asymptotic Bethe ansatz equation was constructed in \cite{Beisert:2005fw}.
It takes the form
\begin{align}
1&=\prod_{j=1}^{K_{4}}\f{x_{4,j}^{+}}{x_{4,j}^{-}}\quad \mbox{(momentum condition)}\,,\\[2mm]
1&=\prod_{j=1}^{K_{2}}\f{u_{1,k}-u_{2,j}+i/2}{u_{1,k}-u_{2,j}-i/2}
	\prod_{j=1}^{K_{4}}\f{1-g^{2}/x_{1,k}x_{4,j}^{+}}{1-g^{2}/x_{1,k}x_{4,j}^{-}}\,,\\[2mm]
1&=\prod_{\mbox{$j=1\atop j\neq k$}}^{K_{2}}\f{u_{2,k}-u_{2,j}-i}{u_{2,k}-u_{2,j}+i}
	\prod_{j=1}^{K_{3}}\f{u_{2,k}-u_{3,j}+i/2}{u_{2,k}-u_{3,j}-i/2}
	\prod_{j=1}^{K_{1}}\f{u_{2,k}-u_{1,j}+i/2}{u_{2,k}-u_{1,j}-i/2}\,,\\[2mm]
1&=\prod_{j=1}^{K_{2}}\f{u_{3,k}-u_{2,j}+i/2}{u_{3,k}-u_{2,j}-i/2}
	\prod_{j=1}^{K_{4}}\f{x_{3,k}-x_{4,j}^{+}}{x_{3,k}-x_{4,j}^{-}}\,,\\[2mm]
\ko{\f{x_{4,k}^{+}}{x_{4,k}^{-}}}^{J}&=\prod_{\mbox{$j=1\atop j\neq k$}}^{K_{4}}\ko{\f{u_{4,k}-u_{4,j}+i}{u_{4,k}-u_{4,j}-i}\,\sig^{2}(u_{4,k},u_{4,j})}\times {}\no\\
	&\quad {}\times \prod_{j=1}^{K_{1}}\f{1-g^{2}/x_{4,k}^{-}x_{1,j}}{1-g^{2}/x_{4,k}^{+}x_{1,j}}
	\prod_{j=1}^{K_{3}}\f{x_{4,k}^{-}-x_{3,j}}{x_{4,k}^{+}-x_{3,j}}
	\prod_{j=1}^{K_{7}}\f{1-g^{2}/x_{4,k}^{-}x_{7,j}}{1-g^{2}/x_{4,k}^{+}x_{7,j}}
	\prod_{j=1}^{K_{5}}\f{x_{4,k}^{-}-x_{5,j}}{x_{4,k}^{+}-x_{5,j}}\,,\label{u4}\\[2mm]
	&{}+\mbox{three more equations for $u_{5}$\,, $u_{6}$ and $u_{7}$\,.}\no
\end{align}
Here the filling numbers $\{K_{\nu}\}_{\nu=1,\dots,7}$ (the excitation numbers of the $\nu$-th node of the diagram) are constrained as $0\leq K_{2}\leq K_{1}+K_{3}\leq K_{4}\geq K_{5}+K_{7}\geq K_{6}\geq 0$\,.
The reference state of the Bethe ansatz equations is chosen to be the ``ferromagnetic'' groundstate of the spin-chain, which corresponds to the BPS operator (\ref{F state}) with 
\begin{equation}
(K_{1},\, K_{2},\, K_{3},\, K_{4},\, K_{5},\, K_{6},\, K_{7})
=(0,\, 0,\, 0,\, 0,\, 0,\, 0,\, 0)\,.
\end{equation}
The rapidity variables $u_{\nu,j}=u(p_{\nu,j})$ are defined through the magnon momentum $p_{\nu,j}$ as
\begin{equation}\label{rapidity}
u(p)=\hf{1}{2}\cot\ko{\hf{p}{2}}\sqrt{1+16g^{2}\sin\ko{\hf{p}{2}}}\,,
\end{equation}
while the spectral parameters $x^{\pm}_{\nu,j}=x^{\pm}(u_{\nu,j})$ and $x_{\nu,j}=x(u_{\nu.j})$ are defined as 
\begin{equation}
x^{\pm}(u)=x\ko{u\pm\hf{i}{2}}\,,\qquad 
x(u)=\hf{1}{2}\ko{u+\sqrt{u^{2}-4g^{2}}}\,.
\end{equation}
The length $L$ of the super spin-chain and the quantum number $J$ in (\ref{u4}) are related as
\begin{equation}
J=L+K_{4}+\hf{1}{2}\ko{K_{1}-K_{3}-K_{5}+K_{7}}\,.
\end{equation}
From string theory perspective, the set of Bethe ansatz equations describes the scattering of string states in the ``decompactifying'' limit $J\to \infty$ so that the gauge-fixed string sigma model becomes a two-dimensional field theory defined on a plane (rather than on a cylinder).

The energy of the spin-chain state is determined by the BPS relation under a centrally-extended $\mathfrak{su}(2|2)$ algebra \cite{Beisert:2005tm}\,.
It is the sum $(\Delta-J={})~\sum_{j=1}^{K_{4}}\ep_{j}$ of the $K_{4}$ main (momentum-carrying) roots with dispersion relation
\begin{equation}
\epsilon_{j}= \sqrt{1+16g^{2}\sin^{2}\ko{\f{p_{4,j}}{2}}}\,,
\label{disp}
\end{equation}
which is actually equivalent to the constraint
\begin{equation}
\bigg(x_{4,j}^{+}+\frac{1}{x_{4,j}^{+}}\bigg) - \bigg(x_{4,j}^{-}+\frac{1}{x_{4,j}^{-}}\bigg)=\frac{i}{g}\,.
\end{equation}
Put it differently, the asymptotically exact formulae for the magnon momenta and energies are given by, in terms of the spectral parameters, 
\begin{align}
p_{4,j}=p(x_{4,j}^{\pm})=\f{1}{i}\ln\bigg(\f{x_{4,j}^{+}}{x_{4,j}^{-}}\bigg)\,,\quad
\ep_{j}=\ep(x_{4,j}^{\pm})=\f{g}{i}\kko{\bigg(x_{4,j}^{+}-\f{1}{x_{4,j}^{+}}\bigg)-\bigg(x_{4,j}^{-}-\f{1}{x_{4,j}^{-}}\bigg)}\,.
\end{align}

The simplest solitonic solution is the elementary (HM) giant magnon ($K_{4}=1$) of \cite{Hofman:2006xt}.
It is the minimum energy state for a given $p:=p_{4}$\,, and thus has a special profile in the spacetime, see the left diagram of Figure\,\ref{fig:GM-GS}; it is just a straight line when viewed from the top.
The energy-spin relation for this state is obtained by taking the HM limit in (\ref{disp}),
\begin{equation}\label{GM}
\Delta-J=\sqrt{1+16g^{2}\sin^{2}\ko{\f{p}{2}}}
\quad \xrightarrow[]{\mbox{\footnotesize $g\to \infty$}} \quad
 4g\left|\sin\ko{\f{p}{2}}\right|\,.
\end{equation}
The physical quantity $p$ in the energy expression (\ref{GM}) corresponds to the geometrical angle between two endpoints of the string when viewed from the top.
In other words, what comes as the argument of sine in (\ref{GM}), that is $p/2$\,, is the ``angular height'' when viewed from the side.
What is more, the energy expression (\ref{GM}) is simply the length of the giant magnon measured with the flat metric on the plane that contains the equatorial circle.

\section{Giant Spinons\label{sec:GS}}

\subsection{Overviews}

One of the aims of the current paper is to identify the states living around the ``top'' of the planar AdS/CFT spectrum.
On the string theory side, we argue that the string state living at the top has vanishing global charges except the energy, {\em i.e.}, the string is at rest.
Likewise, string states living near the top are ``slow-rotating'' strings.
It indicates that the string states are (almost-)circular in shape, wrapping around a great circle of $S^5$\,, namely, they are large-winding states.
The unstablity of slow-rotating circular strings would be related to the fact that the states are located at the top of the spectrum.

\begin{figure}[tbp]
\begin{center}
\vspace{0.5cm}
\includegraphics{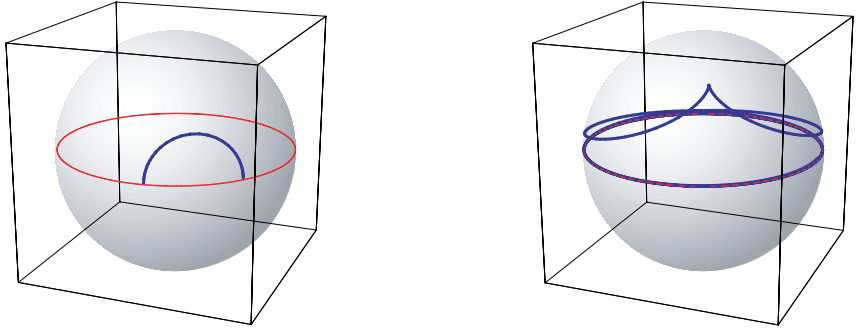}
\vspace{0.3cm}
\caption{\sf A giant magnon (left) and a single-spike string (right).}
\label{fig:GM-GS}
\end{center}
\vspace{0.5cm}
\end{figure}

Now it is easy-to-guess how the string at the top looks like\,; it is an infinitely-wound circular string wrapping around a great circle of $S^5$\,.
Below we claim that it is such a hoop-like string that corresponds to the antiferromagnetic (AF) state of the AdS/CFT spin-chain.
String states corresponding to spinon excitations are then obtained by exciting the hoop string state.
Actually we know a nice candidate for such a state\, \----\, the so-called single-spike string found in \cite{Ishizeki:2007we,Mosaffa:2007ty} as the first example of large-winding string and further studied in \cite{Hayashi:2007bq,Ishizeki:2007kh,Abbott:2008yp,Ahn:2008sk} and others.
The diagram is shown in the right of Figure\,\ref{fig:GM-GS}.

\paragraph{}
On the gauge theory side, as the AF state living at the top of the spectrum, we consider a spin-neutral state with the largest possible energy (anomalous dimension) for a given length $L$\,.
It falls into a category of operators of the form, schematically,\footnote{Here we do not mean the state is factorised into a product of $\cS$\,.
Each $\cS$ is decomposed, and the constituent excitations mix with those from other $\cS$'s under renormalisation.}
\begin{align}\label{AF state}
\mathcal O_{\rm AF}&\sim \tr\left( \mathcal S^{L/2} \right)+\mbox{permutations}\no\\
&\eq\tr\left( \cS(\{x_{1}^{\pm}\})\cS(\{x_{2}^{\pm}\}) \dots \cS(\{x_{L/2}^{\pm}\}) \right)+\mbox{permutations}\,,
\end{align} 
where each $\mathcal S(\{x_{j}^{\pm}\})$ ($j=1,\dots,\hf{L}{2}$) stands for particular spin-neutral composites of $\mathcal N=4$ fields specified by spectral parameters $\{x_{j}^{\pm}\}=( x_{1,j}^{\pm},\dots,x_{7,j}^{\pm})$\,.
The near-AF states will be then represented schematically as
\begin{equation}\label{spinons}
\mathcal O_{\rm spi}\sim \tr\left( \mathcal X_{1}\dots\mathcal X_{M}\mathcal S^{L/2-M} \right)+\mbox{permutations}\,,
\end{equation} 
where $M\ll L$\,, and $\mathcal X_{j}$ are arbitraly elementary fields of $\mathcal N=4$ SYM as before.
Below we will show that a specific configuration of Bethe roots, or the spectral parameters, indeed reproduces the energy-spin relation (dispersion relation) for a single-spike string of \cite{Ishizeki:2007we,Mosaffa:2007ty}.

\subsection{The Antiferromagnetic State}

Let us write the Dynkin labels of the bosonic subalgebras $\mathfrak{su}(4)$ and $\mathfrak{su}(2,2)$ of $\cN=4$ global symmetry $\mathfrak{psu}(2,2|4)$ as $[q_{1},p,q_{2}]$ and $[s_{1},r,s_{2}]$\,, respectively.
They are expressed in term of the filling numbers $\{ K_{\nu} \}$\,, the length $L$ and the anomalous dimension $\gamma$ as \cite{Beisert:2005fw}
\begin{align}
\hspace{-.25cm}[q_{1},p,q_{2}]&=[-(K_{1}+K_{3})+K_{4}\,,\,L+(K_{3}+K_{5})-2K_{4}\,,\,-(K_{7}+K_{5})+K_{4}]\,,\label{charge S}\\
\hspace{-.25cm}[s_{1},r,s_{2}]&=[(K_{1}+K_{3})-2K_{2}\,,\,-L+(K_{2}+K_{6})-(K_{3}+K_{5})-\gamma\,,\,(K_{7}+K_{5})-2K_{6}]\,.\label{charge AdS}
\end{align}
As discussed the AF state we propose has vanishing global charges except the energy.
This condition constraints the filling numbers as 
\begin{equation}
K_{1}+K_{3}=2K_{2}=K_{4}=2K_{6}=K_{7}+K_{5}\qquad 
{\rm and}\qquad
L=2K_{4}-(K_{3}+K_{5})\,.
\end{equation}
Let us now set $K_{4}=:2M$\,, and also set $K_{1}=K_{7}=:K$ since for the AF state the filling numbers must be symmetric about the central node $\nu=4$\,.
Then the AF state falls into a class of states specified by the following fillings
\begin{equation}\label{fillings2}
(L\,;\, K_{1},\, K_{2},\, K_{3},\, K_{4},\, K_{5},\, K_{6},\, K_{7})
=(2K\,;\, K,\, M,\, 2M-K,\, 2M,\, 2M-K,\, M,\, K)\,.
\end{equation}
The energy of the state is given by $\Delta-J=\sum_{j=1}^{K_{4}}\ep_{j}$ with $\ep_{j}$ as in (\ref{disp}).
It is independent of $K$\,, or the length of the state, which we fix to (formally) define a state with the largest possible energy.\footnote{Note also that one can always tune the value of $K$ between $0$ and $2M$ via the so-called dynamic duality transformations $x_{3}\mapsto g^{2}/x_{1}$ and $x_{5}\mapsto g^{2}/x_{7}$ \cite{Beisert:2005fw}.}
In order to maximise the energy for the fixed number of momentum-carrying magnons $K_{4}=2M$\,, we take the maximal number of magnons such that the Bethe roots are all real.
In the well-known case of the AF vacuum of the Heisenberg (XXX$_{1/2}$) spin-chain, there is one excitation per each mode number (corresponding to each branch of the log) and all available levels are filled.
In the current case of the AdS/CFT spin-chain, we claim that the AF state is made up of the same number $m~(\gg 1)$ main $(\nu=4)$ excitations per each mode number $n$\,.
Here all available levels $n=-N_{0},\dots,N_{0}$ are filled where $N_{0}$ is an integer closest to $2g$\,, and we label the $m$ roots at each mode $n$ as $u_{n,1},\dots,u_{n,m}$\,.
In other words, $2M\eq m\cdot 2N_{0}$ roots are equally distributed into $2N_{0}$ adjacent points $u=-N_{0},\dots,N_{0}$ on the real axis.
Using physical terminology we can call this state the Dirac sea of magnons.

\begin{figure}[tbp]
\begin{center}
\vspace{-0.5cm}
\includegraphics{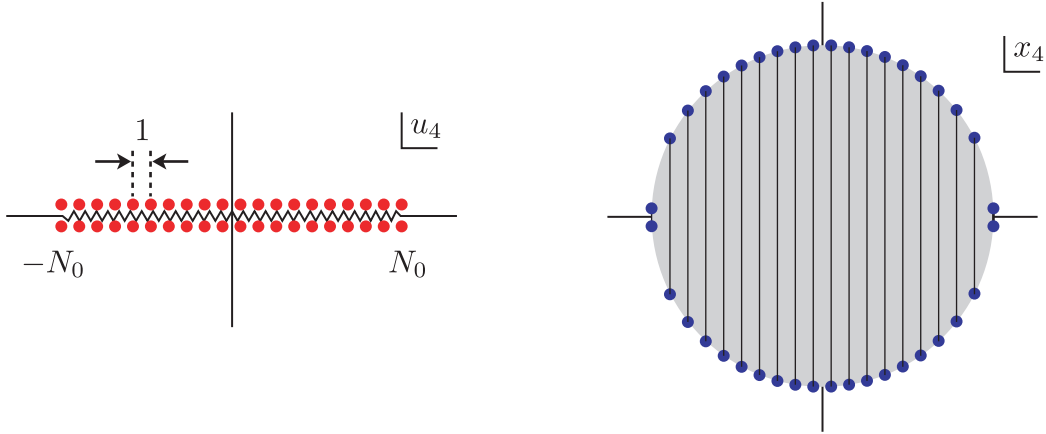}
\vspace{0.5cm}
\caption{\sf The antiferromagnetic state of AdS/CFT in the infinite-winding limit. 
(Left\,:\,rapidity plane, Right\,:\,spectral parameter plane.)}
\label{fig:AF_ground}
\end{center}
\end{figure}

\paragraph{}
In order to make contact with classical string theory, now we take the strong coupling limit and the thermodynamic limit in addition to the infinite-winding limit (\ref{LW limit}).
As in the infinite-spin limit (\ref{HM limit}) of HM, in the infinite-winding limit (\ref{LW limit}) of our concern also, solutions of integral equations that are the thermodynamic limit of the Bethe ansatz equations are described only by condensate cuts in the rapidity plane \cite{Minahan:2006bd}.
This is because the infinite-winding state in question is essentially a superposition of infinite number of infinite-spin states, where each spin can be both positive infinite and negative infinite such that all sum up to yield a finite net spin.
It is consistent with the fact that, in string theory, the profile of the spiky string is given by trigonometric functions.
Notice also that in the finite-gap language the winding number $m$ is interpreted as the period of a particular cycle about condensate cuts.

In the infinite-winding limit (\ref{LW limit}), for each condensate labeled by $n$\,, the constituent roots $u=u_{n,1},\dots,u_{n,m}$ approaches the corresponding centre point on the real axis, thus shrinking to a singular, zero-length cut at $u=n$\,.
See Figure\,\ref{fig:AF_ground} for the diagram.
Denoting $p_{n,k}:=p(u_{n,k})$\,, the condition (\ref{rapidity}) now translates to
\begin{equation}\label{momentum}
\cos\ko{\f{p_{n,k}}{2}}=\f{n}{N_{0}}\qquad 
\mbox{for}\quad n=-N_{0},\, \dots,\, +N_{0}\quad 
\mbox{and}\quad k=1,\, \dots,\, m
\end{equation}
in this limit.
For such a state, the energy above the ferromagnetic vacuum (or the light-cone energy) in the strong coupling limit is evaluated as
\begin{align}
\Delta-J&=\lim_{g\to \infty}\kko{4g\sum_{j=1}^{2M}\sin\ko{\f{p_{j}}{2}}}
=\lim_{g,\,N_{0}\to \infty}\kko{4g\sum_{n=-N_{0}}^{N_{0}}\sum_{k=1}^{m}\sin\ko{\f{p_{n,k}}{2}}}\no\\[2mm]
&=\lim_{N_{0}\to \infty}\kko{8g\cdot mN_{0}\int_{0}^{1}dy\,\sqrt{1-y^{2}}}=m\sqrt{\lam}\cdot \lim_{N_{0}\to \infty}N_{0}/2\,.\label{hoop}
\end{align}
Comparing this with the energy of a $m$-times wound hoop-string \cite{Okamura:2006zv}, the former is just $N_{0}/2$ copies of the latter.
In the next subsection we will see more general example of the matching of the spectra which includes this AF/hoop case as a special case.

This infinite factor $N_{0}$ appearing here is understood as follows.
As discussed in \cite{Hayashi:2007bq}, the hoop-string is obtained from a point-like, BPS string by swapping the time ($\tau$) and space ($\sig$) worldsheet variables in the sphere part of the string profile while the AdS part unchanged.
This is a general feature about the relation between large-winding string states and the corresponding large-spin states in the conformal gauge \cite{Hayashi:2007bq}\,; writing the six-vector in the sphere subspace as $\vec X(\tau,\sig)$\,, naively it follows that $\vec X_{\rm hoop}(\tau,\sig)=\vec X_{\rm BPS}(\sig,\tau)$ and $\vec X_{\rm spike}(\tau,\sig)=\vec X_{\rm GM}(\sig,\tau)$\,, {\em etc.}.
We must, however, properly set the ranges of the worldsheet variables. 
After the $\tau\leftrightarrow \sigma$ flip, the original time variable plays the role of the space variable and {\em vice versa}, therefore the definition ranges of the variables need to be both infinite.
A convenient way to realise it is to define the rescaled worldsheet variables $(\tau,\sig)=\kappa (\tau',\sig')$\,, where the primed variables are the original ones defined such that $-\infty<\tau'<\infty$ and $-\pi<\sig'<\pi$\,, then take $\kappa\to \infty$\,.
As the result, the space variable after the $\tau\leftrightarrow \sigma$ flip takes values in a doubly infinite region, {\em i.e.}, there are infinite number of copies of string profile, each defined in the infinite range $(-\infty,\infty)$\,.
Thus the infinity ($N_{0}\to \infty$) in (\ref{hoop}) is associated with the infiniteness of worldsheet time before the $\tau\leftrightarrow \sigma$ flip.

\paragraph{}
So far we have mostly cared about the configuration of main roots $(\nu=4)$\,.
As for the other roots ($\nu\neq 4$), we may assume that in the thermodynamic limit they form kind of boundstates called stacks \cite{Beisert:2005di} as is usual for other maximally filled cases \cite{Rej:2007vm}.
For the highest energy state, we may assume $u_{2,j}=u_{6,j}$ for $j=1,\dots,M$ by symmetry, with each neighbouring set of roots $u_{1,j}$ and $u_{2,k}$ attract each other to form a stack.
With appropriate ordering of the roots, the stack configuration is expressed, denoting either of $u_{1}$ or $u_{3}$ as $v$\,, as $v_{2k-1}\approx u_{2,k}+\hf{i}{2}$ and $v_{2k}\approx u_{2,k}-\hf{i}{2}$ ($k=1,\dots,M$)\,, where the symbol ``$\approx$'' indicates equality up to $\cO\big(\hf{1}{L}\big)$ correction, which in our thermodynamic case can be treated as ``$=$''.

In summary, the main roots $u_{4,j}^{\pm}$ condensed on the real axis of the rapidity plane as in (\ref{momentum}) together with the other six auxiliary roots $u_{\nu\neq 4,j}^{\pm}$ forming symmetric stacks as discussed above, in total, form the spin-neutral composites $\cS(\{ x_{j}^{\pm} \})$ appearing in (\ref{AF state}).

\subsection{Special Excited States \---- ``Giant Spinons''}

\begin{figure}[tbp]
\begin{center}
\vspace{-0.5cm}
\includegraphics{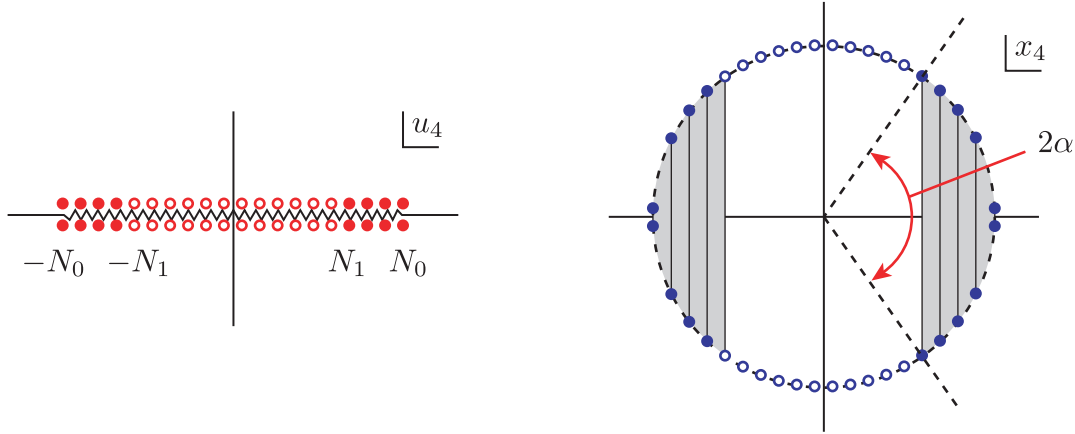}
\vspace{0.5cm}
\caption{\sf An excited state in the infinite-winding limit.
Each tiny filled $(\bullet)$ and open ($\circ$) circle contains $m+1$ and $m$ roots, respectively.
(Left\,:\,rapidity plane, Right\,:\,spectral parameter plane.)}
\label{fig:spinons}
\end{center}
\end{figure}

Let us now go on to the excited states.
In solid state physics such as the Heisenberg spin-chain, excitations above the AF vacuum are known as ``spinons''.
We argue that certain collective modes of macroscopic (infinite) number of magnons correspond to the spiky string states in string theory.
Our proposal for such root configurations is shown in Figure\,\ref{fig:spinons}.
It is again described as a filled sea of magnons, but this time with
\begin{equation}\label{N_1}
\mbox{\# of roots at mode $n$}=
\left\{\begin{array}{cl}
m+1	& (-N_{0}\leq n < -N_{1}\,,\, +N_{1} < n\leq N_{0}) \\
m	& (-N_{1}\leq n\leq +N_{1})
\end{array}\right.\,,
\end{equation}
where $N_{1}$ is an integer between $0$ and $N_{0}$\,.
For later convenience let us introduce the notation $\cB [-N_{1},N_{1}|m]$ to refer to the state with excitation mode (\ref{N_1}).\footnote{With this notation the AF vacuum state discussed above is represented as $\cB [-N_{0},N_{0}|m]$ or $\cB [0,0|m-1]$\,.\label{def:cB}}
The energy above the ferromagnetic vacuum is then obtained in the same manner as before, yielding
\begin{align}\label{disp rel}
\Delta-J&=\lim_{g\to \infty}\kko{4g\sum_{j=1}^{2M}\sin\ko{\f{p_{j}}{2}}}
=\lim_{g,\,N_{0}\to \infty}\kko{8gN_{0}\ko{m\int_{0}^{1}+\int_{\cos\al}^{1}}dy\,\sqrt{1-y^{2}}}\no\\[2mm]
&=\f{\sqrt{\lam}}{\pi}\ko{m\pi +2\al-\sin(2\al)}\cdot \lim_{N_{0}\to \infty}N_{0}/2\,.
\end{align}
Comparing this expression with the known ``funny'' energy-spin relation for a single-spin single-spike with angular height $\bar\theta$ and infinite winding angle $\Delta\varphi$ \cite{Ishizeki:2007we,Mosaffa:2007ty},
\begin{equation}
E-\f{\sqrt{\lam}}{2\pi}\,\Delta\varphi =\f{\sqrt{\lam}}{\pi}\,\bar \theta\,,\quad
J=\f{\sqrt{\lam}}{\pi}\,\sin\bar\theta
\quad \Rightarrow\quad 
E-J=\f{\sqrt{\lam}}{\pi}\ko{\f{\Delta\varphi}{2}+\bar\theta-\sin\bar\theta}\,,
\end{equation}
they precisely match under identification $2\al\eq \bar \theta$ provided that the condition $\Delta\varphi=2\pi m$ holds.
The interpretation of the infinite factor $N_{0}/2$ is the same as before (this time it counts the number of spikes).

Since for given $N_{0}$ or the coupling $g$\,, the total winding angle $\Delta\varphi_{\rm tot.}=\pi m N_{0}$ for the superposed $N_{0}/2$ spikes only depends on the integer $m$\,, changing the AF vacuum $\cB[-N_{0},N_{0}|m]$ to $\cB[-N_{1},N_{1}|m]$ ($0\leq N_{1}\leq N_{0}$) does not affect the winding angle $\Delta\varphi_{\rm tot.}$\,.
However, as mentioned in Footnote \ref{def:cB}, when the number of excitations above $\cB[-N_{0},N_{0}|m]$ increase and eventually $N_{1}$ hits $0$\,, there emerges a ``new'' AF vacuum $\cB[0,0|m]\equiv \cB[-N_{0},N_{0}|m+1]$\,, changing the winding angle to $\Delta\varphi=\pi (m+1) N_{0}$\,.
Note that this new vacuum is still physically equivalent to the old one in that $m$ is set to be infinite in our construction of the AF vacuum.

It is instructive to understand this situation from the string theory point of view.
Let us start with a hoop string ($\simeq \cB[-N_{0},N_{0}|m]$) corresponding to a series of spikes with each spike having zero angular height, $\bar\theta=0$\,.
As $\bar\theta$ increases, the string becomes extended, and when it gets enough slack to wrap around the great circle once more to recover the hoop shape ($\simeq \cB[-N_{0},N_{0}|m+1]$), the winding number changes by $+1$ from the original hoop.

Again, we face a kind of fancy situation here\,; the geometrical angle $\bar\theta$ in string spacetime is directly identified with $2\al$ (satisfying $\cos\al=N_{1}/N_{0}$) in the spectral parameter plane.
Also, the infinite winding angle $\Delta\varphi_{\rm tot.}$ is identified with ($1/2$ of) the momentum of background sea of magnons, that is the AF state of the spin-chain.
To see this, one only needs to integrate the momentum (\ref{momentum}) as
\begin{equation}
P_{\rm AF}=\sum_{n=-N_{0}}^{N_{0}}\sum_{k=1}^{m}p_{n,k}
=m\sum_{n=-N_{0}}^{N_{0}}2\arccos\ko{\f{n}{N_{0}}}
=2\pi m\cdot N_{0}
\qquad (N_{0}\sim\sqrt{\lam}\to \infty)\,.
\end{equation}
Similarly, one can calculate the momentum shift caused by the spinon excitation as $P_{\rm spi}-P_{\rm AF}=2\pi \ko{1-\cos\al} N_{0}$\,.
These features can be compared to the ordinary giant magnon case \cite{Hofman:2006xt}, in which the projection of the string profile onto the ``equatorial plane'' is directly identified with a straight stick in the LLM plane \cite{Lin:2004nb} whose endpoints being located on the ``equatorial circle'', see the comments at the end of the previous section.
It can be further identified with the finite-gap description in string theory, or the Bethe string configuration in gauge theory.
It is also interesting to notice that the energy (\ref{disp rel}) essentially represents the area of the shaded region in the complex spectral parameter plane in Figure\,\ref{fig:spinons}.

One more remark is that the charges $\Delta$ and $J$ are not obtained separately on the spin-chain side in contrast to the string theory.
As shown in \cite{Beisert:2005fw}, these spin-chain charges can be actually expressed by the filling numbers $\{ K_{\nu} \}$\,, the length $L$ and the anomalous dimension $\ga$ through the relations $\Delta=-\hf{1}{2}s_{1}-r-\hf{1}{2}s_{2}$ and $J=\hf{1}{2}q_{1}+p+\hf{1}{2}q_{2}$ (see (\ref{charge S}-\ref{charge AdS})).
However, when combined together to give the difference of these charges, all auxiliary terms drop off, leaving the simple expression $\Delta-J=K_{4}+\ga$\,.
In our case, it leads to (\ref{disp rel}) in terms of $m$\,, $N_{0}$ and $N_{1}$ (or $\al$).
Thus we are comparing the spectra of string and gauge counterparts in the form of $\{ \Delta \mbox{(or $E$)} - J \}_{\rm states}$ in this decompactifying limit.

\subsection{Classical Scattering of Giant Spinons}

\begin{figure}[tbp]
\begin{center}
\vspace{-0.5cm}
\includegraphics{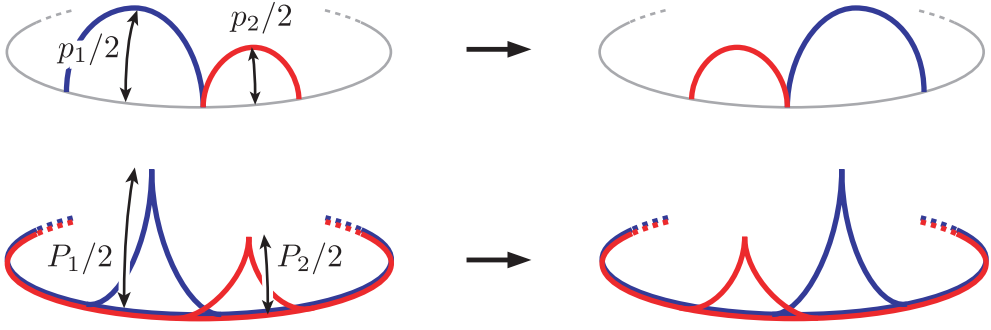}
\vspace{0.5cm}
\caption{\sf Scattering of giant-magnons/single-spikes.}
\label{fig:scatter}
\end{center}
\end{figure}

Finally, we comment on implications of the paper \cite{Ishizeki:2007kh}.
In the paper, a scattering state of two single-spikes were constructed by the so-called dressing method (see also \cite{dressing}), and using the solution, the scattering phase-shift for two single-spikes were determined as a function of angular heights $\bar\theta_{j}=:P_{j}/2$ ($j=1,2$)\,.
Remarkably, the phase-shift $\Th_{\rm spi}(P_{1},P_{2})$ agreed with that for giant magnons $\Th_{\rm mag}(p_{1},p_{2})$ (here $p_{j}/2$ are the angular heights of giant magnons), up to non-logarithmic (gauge-dependent) terms.
See Figure\,\ref{fig:scatter} for the spacetime picture of how the magnons/spinons scatter.
This is a kind of puzzle since the giant magnons and single-spikes have totally different $(\tau,\sig)$-dependence, and so there seems no reason the classical formula $\Delta T=\pa\Theta/\pa \ep$\,, relating the time delay $\Delta T$\,, the soliton energy $\ep$ and the scattering phase-shift $\Theta$\,, gives the same result for the two kinds of soliton scatterings.
Below we will give a possible explanation for the occurrence of the agreement by using our Bethe string picture.

A scattering state of two spikes with angular heights $P_{1}/2$ and $P_{2}/2$ are described, using the notation introduced before, as $\cB[-N_{1},N_{1}|m_{1}]\cup \cB[-N_{2},N_{2}|m_{2}]$\,, where $\cos (P_{j}/2)=N_{j}/N_{0}$ and $m_{j}\gg 1$ ($j=1,2$).
Below we will abbreviate $\cB[-N_{j},N_{j}|m_{j}]$ as $\cB_{j}$ for short.
Recall that, in the strong coupling limit, the scattering phase-shift for two elementary magnons with spectral parameters $x_{1}^{\pm}$ and $x_{2}^{\pm}$ becomes
\begin{equation}
-i\ln S(x_{1}^{\pm},x_{2}^{\pm})
=\theta(x_{1}^{\pm},x_{2}^{\pm})
=2g\left[k(x_{1}^{+},x_{2}^{+})+k(x_{1}^{-},x_{2}^{-}) -k(x_{1}^{+},x_{2}^{-})-k(x_{1}^{-},x_{2}^{+})\right]\,.
\label{factorizedAFS}
\end{equation}
In the strong coupling limit, the function $k(x,y)$ can be expanded in powers of $1/g$ as $k(x,y)=\sum_{n}g^{-n}k_{n}(x,y)$\,.
The tree-level contribution is given by \cite{Arutyunov:2004vx}
\begin{equation}
k_{0}(x,y)=-\left[\left(x+\frac{1}{x}\right)-\left(y+\frac{1}{y}\right)\right]\ln\left(1-\frac{1}{xy}\right)
\label{elemetarychi0}\,.
\end{equation}
When two giant magnons with spectral parameters $x_{j}^{\pm}=e^{\pm ip_{j}/2}$ ($j=1,2$) scatter, the resulting phase-shift becomes 
\begin{equation}
\hspace{-.0mm}
\theta(x_{1}^{\pm},x_{2}^{\pm})\Big|_{x_{j}^{\pm}=e^{\pm ip_{j}/2}}
=\kko{\cos\ko{\f{p_{1}}{2}}-\cos\ko{\f{p_{2}}{2}}}\ln\kko{\f{1-\cos\ko{\hf{p_{1}-p_{2}}{2}}}{1-\cos\ko{\hf{p_{1}+p_{2}}{2}}}}
=: \Theta_{\rm mag}(p_{1},p_{2})\,.
\end{equation}
Obviously, this function satisfies $\Theta_{\rm mag}(p,q)+\Theta_{\rm mag}(p,-q)=0$ for $p\neq q$\,.
It is also easy to check that $\Theta_{\rm mag}(p,p)=\Theta_{\rm mag}(p,-p)=0$\,.

On the other hand, a scattering phase-shift for two giant spinons $\cB_{1}$ and $\cB_{2}$ can be obtained by virtue of integrability as factorised scatterings of infinite number of elementary magnon,
\begin{equation}\label{S_{1,2}}
\Theta_{\rm spi}(P_{1},P_{2})=-i\ln S_{\cB_{1},\cB_{2}}
\quad \mbox{with}\quad 
S_{\cB_{1},\cB_{2}}=\prod_{j_{1}\in \cB_{1}}\prod_{j_{2}\in \cB_{2}}S(x_{j_{1}}^{\pm},x_{j_{2}}^{\pm})\,.
\end{equation}
A nontrivial feature observed in \cite{Ishizeki:2007kh} is that the function $\Theta_{\rm mag}(p_{1},p_{2})$ is precisely the same as $\Theta_{\rm spi}(P_{1},P_{2})$\,.
Notice that, naively, $S_{\cB_{1},\cB_{2}}$ in (\ref{S_{1,2}}) reduces to unity since both sets $\cB_{j}$ of real roots are symmetric about the imaginary axis.
However, this can be thought of rather an artifact due to the singular nature of the spinon configurations, and it can be cured by taking into account the quantum boundary effect of the configuration within $\cO(g^{-1})$\,.
As a convenient choice, we may consider the scattering of $\cB'_{j}\eq \cB[-N_{j},N_{j}+1|m]$ instead of $\cB_{j}=\cB[-N_{j},N_{j}|m]$\,.
Accordingly, $\cos P'_{j}/2=(N_{j}+1)/N_{0}$\,.
This subtle change does not affect the energy-spin relations nor the angular heights to the first order in the strong-coupling (large-$N_{0}$) expansion, but do the phase-shift as
\begin{align}
\Theta_{\rm spi}(P_{1},P_{2})&=\sum_{j_{1}\in \cB'_{1}}\sum_{j_{2}\in \cB'_{2}}\Theta_{\rm mag}(p_{j_{1}},p_{j_{2}})\no\\
&=\sum_{j_{1}\in \cB_{1}}\sum_{j_{2}\in \cB_{2}}\Theta_{\rm mag}(p_{j_{1}},p_{j_{2}})
+\sum_{j_{2}\in \cB_{2}}\Theta_{\rm mag}(P'_{1},p_{j_{2}})
+\sum_{j_{1}\in \cB_{1}}\Theta_{\rm mag}(p_{j_{1}},P'_{2})+{}\no\\
&\hphantom{{}={}}{}+\Theta_{\rm mag}(P'_{1},P'_{2})\,.
\end{align}
Since $\cB_{j}$ are both symmetric about the imaginary axis, all terms except the last one vanish, leading to $\Theta_{\rm mag}(P'_{1},P'_{2})=\Theta_{\rm spi}(P_{1},P_{2})$ to the first order as expected.
The result is the same if we consider the scattering of $\cB[-N_{j}+\delta_{L,j},N_{j}+\delta_{R,j}|m]$ as long as the nonzero fluctuations $|\delta_{j}|$ are within $\cO(g^{-1})$\,.

\paragraph{}
In the discussion above, we have not included the non-universal, gauge-dependent term that appears when the computation is performed in the string frame.
Explicitly, in the dyonically charged (boundstate) case \cite{scatt:boundstate}, it takes the form $(\ep{}^{(2)}-J{}^{(2)}_{2})P^{(1)}$ in the expression of the scattering phase-shift, where $\ep^{(2)}=E^{(2)}-J{}^{(2)}_{1}$ and $J^{(2)}_{2}$ are the light-cone energy and the second spin of the scattering particle 2 respectively, and $P^{(1)}$ is the momentum of the particle 1.
This sort of term can be accounted for by taking into account the different effective length of the excitation on the both sides of the AdS/CFT correspondence \cite{Hofman:2006xt,scatt:boundstate}.%
\footnote{On the string theory side, we work in the conformal gauge, so that the density of $E$ is constant, whereas on the spin-chain side a unit length is assigned to each $\mathfrak{su}(2)$ site $\cZ$ or $\cW$\,.
Hence when there are $J_{1}$ $\cZ$s and $J_{2}$ $\cW$s in the spin-chain, we have $\Delta \ell_{\text{spin-chain}}=\int d(J_{1}+J_{2})=\int dE-\int d\kko{(E-J_{1})-J_{2}}=\Delta x_{\rm string}-(\ep-J_{2})\,.$
By exponentiating it, $S_{\rm string}=S_{\text{spin-chain}}\,e^{i(\Delta x_{\rm string}-\Delta \ell_{\text{spin-chain}})^{(2)}P^{(1)}}=S_{\text{spin-chain}}\,e^{i(\ep^{(2)}-J^{(2)}_{2})P^{(1)}}$ as expected.}

As shown in \cite{Ishizeki:2007kh}, the non-universal term for a single-spin single-spike string scattering is given by $\tilde \ep^{(1)}\tilde J^{(2)}/2g$\,, where $\tilde \ep=4g\bar\theta$ and $\tilde J=4g\sin\bar\theta$\,.
One can easily check that this term is also reproduced in the same manner as above, namely by integrating all the contributions of factorised magnon scattering $\ep_{j_{2}}^{(2)}p_{j_{1}}^{(1)}$ over $j_{1,2}\in\cB_{1,2}$\,.
Explicitly, it is given by $4gP_{1}'\sin(P_{2}'/2)$ with $P_{j}'$ and $\theta_{j}$ defined as before.

\section{Conclusions and Discussions\label{sec:C&D}}

In this paper, we explored around the AF state of the AdS${}_{5}$/CFT${}_{4}$ correspondence.
By taking the novel infinite-winding limit (\ref{LW limit}) together with the strong coupling limit $g\to \infty$ as well as the thermodynamic limit $L\to \infty$\,, we provided the first example of dual pairs in the large-winding sector of the AdS/CFT.

The AF vacuum state of the spin-chain description of the AdS/CFT system is given by a filled Dirac sea of magnons.
Explicitly, it is described by infinite number $2M$ of momentum-carrying ($\nu= 4$) roots equally distributed into $2N_{0}$ adjacent points $u=-N_{0},\dots,N_{0}$ on the real axis of the complex rapidity plane, where each portion $m=M/N_{0}$ is infinite, and $N_{0}\sim g$ is also very large if we take the strong coupling limit.
As for the other auxiliary roots ($\nu\neq 4$), we assume that they form stacks.
The energy of the AF vacuum measured from the ferromagnetic vacuum agreed with the energy of the hoop-string, which is the purely winding string state with no spins.

Then we studied the low-lying excitations, which are holes in the filled Dirac sea.
The specific excitation mode which we called the giant spinon reproduced the energy-spin relation for a spike string solution on the string theory side.
We claimed that the special collective excitation mode $\cB_{1}\eq\cB[-N_{1},N_{1}|m]$ defined in (\ref{N_1}) (see Figure\,\ref{fig:spinons}) corresponds to the spike string by showing that $(i)$ $\cB_{1}$ reproduces the energy-spin relation for a spike string, and that $(ii)$ the identification allowed us to give, from the spin-chain viewpoint, a possible explanation for the phenomenon observed in \cite{Ishizeki:2007kh}, that is the scattering phase-shift for two giant magnons agrees with that for two single-spikes.
These evidences lead us to conjecture that our giant spinon configuration of Bethe roots is the long-sought gauge theory dual of a classical single-spike string.

\begin{table}[tbp]
\vspace{-.8cm}
\caption{\sf The AdS/CFT spectrum from the spin-chain perspective.}
\begin{center}
\vspace{-.0cm}
{\small 
\begin{tabular}{|l||l|l|l|} \hline
	& String state 
		& SYM operator 
			& Spin-chain\\ \hline\hline
\multirow{3}{*}{Large-Spin}	
	& Point-particle
		& $\tr(\mathcal Z^{L})$ 
			& Ferro. groundstate\\ \cline{2-4}
{}
	& Tiny string
		& $\tr(\mathcal X^{M}\mathcal Z^{L-M})+\mbox{\rm perm.}$\,,~$M\ll L$
			& Magnon excitations\\ \cline{2-4}
{}
	& Large string
		& $\tr(\mathcal X^{M}\mathcal Z^{L-M})+\mbox{\rm perm.}$\,,~$M\sim\mathcal O(L)$
			& Classical spin-waves\\ \hline
\multirow{2}{*}{Large-Winding}	
	& Spiky string
		& $\tr(\cX^{M}\cS^{L/2-M})+\mbox{\rm perm.}$\,,~$M\ll L$ 
			& Spinon excitations\\ \cline{2-4}
{}
	& Hoop string
		& $\tr(\cS^{L/2})+\mbox{\rm perm.}$
			& Antiferro. groundstate\\ \hline
\end{tabular}
}
\end{center}
\label{tab:spectrum}
\vspace{.0cm}
\end{table}

We summarised our knowledge regarding the AdS/CFT spectrum from the spin-chain perspective in Table\,\ref{tab:spectrum}.
In a nutshell, contribution of this paper is that we added the last two lines, ``AF/hoop'' and ``near-AF/spikes'', by defining the special infinite-winding limit (\ref{LW limit}).
These states live very far from BPS.

\paragraph{}
In closing, we comment on some possible applications of our results.
Generalising our interpretation of single-spin solutions to multi-spin cases would be the first thing to do, just like all the stories about the giant magnons \cite{boundstate,scatt:boundstate}.
This means to construct a duality map between multi-spin single-spikes \cite{Ishizeki:2007we} and some yet-to-be discovered configuration of Bethe roots, possibly a kind of boundstates (see \cite{Hayashi:2007bq} for related viewpoints).

Spinon excitations that are not ``giant'', in the sense that holes are not such well-organised as (\ref{N_1}), would also be interesting objects to investigate.
Computing the quantum fluctuations around spinon states would be useful in order to further test our proposal.
The non-linear integral equation approach developed in the context of $\cN=4$ SYM in {\em e.g.} \cite{NLIE} would be useful.

Investigating ``finite-size corrections'' for the large-winding states would also be possible.
Such corrections to single-spikes were worked out in \cite{Abbott:2008yp,Ahn:2008sk}, and it is interesting to reproduce the results from the corresponding Bethe string picture (especially in a finite-gap language).
``Large-$m$(winding)'' expansion' of the AdS/CFT spectrum around the AF state may be possible, albeit complicated.
The instability of the large-winding string states explicitly shown in \cite{Abbott:2008yp} is also worthy of note.
Reconfiguration of Bethe roots in the AF state may reproduce the decay rate obtained in the paper.

Furthermore, generalisation to the AdS${}_{4}$/CFT${}_{3}$ duality \cite{Aharony:2008ug} is also important.
The integrability of the system is intensively studied \cite{AdS4CFT3}, while at the level of concrete solutions, giant magnons as well as single-spike strings on $AdS_{4}\times {\mathbb{CP}}^{3}$ background are known \cite{AdS4CFT3:giants}, so that one should be able to generalise the correspondence proposed in the curret paper to the case of the novel AdS${}_{4}$/CFT${}_{3}$\,.

\subsubsection*{Acknowledgments}

We would like to thank Yosuke Imamura, Nick Dorey, Radu Roiban, Ryo Suzuki and Arkady Tseytlin for valuable comments on the manuscript.


\providecommand{\href}[2]{#2}\begingroup\raggedright\endgroup


\begin{thebibliography}{10}

\bibitem{AdSCFT}
J.~M. Maldacena, ``{The large $N$ limit of superconformal field theories and
  supergravity},'' {\em Adv. Theor. Math. Phys.} {\bf 2} (1998) 231--252,
\href{http://arXiv.org/abs/hep-th/9711200}{{\tt hep-th/9711200}}.
~{\small $\bullet$}~
S.~S. Gubser, I.~R. Klebanov, and A.~M. Polyakov, ``{Gauge theory correlators
  from non-critical string theory},'' {\em Phys. Lett.} {\bf B428} (1998)
  105--114,
\href{http://arXiv.org/abs/hep-th/9802109}{{\tt hep-th/9802109}}.
~{\small $\bullet$}~
E.~Witten, ``{Anti-de Sitter space and holography},'' {\em Adv. Theor. Math.
  Phys.} {\bf 2} (1998) 253--291,
\href{http://arXiv.org/abs/hep-th/9802150}{{\tt hep-th/9802150}}.

\bibitem{Minahan:2002ve}
J.~A. Minahan and K.~Zarembo, ``{The Bethe-ansatz for ${\mathcal N}=4$ super
  Yang-Mills},'' {\em JHEP} {\bf 03} (2003) 013,
\href{http://arXiv.org/abs/hep-th/0212208}{{\tt hep-th/0212208}}.

\bibitem{Bena:2003wd}
I.~Bena, J.~Polchinski, and R.~Roiban, ``{Hidden symmetries of the
  $AdS_{5}\times S^{5}$ superstring},'' {\em Phys. Rev.} {\bf D69} (2004)
  046002,
\href{http://arXiv.org/abs/hep-th/0305116}{{\tt hep-th/0305116}}.

\bibitem{Beisert:2003tq}
N.~Beisert, C.~Kristjansen, and M.~Staudacher, ``{The dilatation operator of
  ${\mathcal N}=4$ super Yang-Mills theory},'' {\em Nucl. Phys.} {\bf B664}
  (2003) 131--184,
\href{http://arXiv.org/abs/hep-th/0303060}{{\tt hep-th/0303060}}.

\bibitem{Beisert:2003yb}
N.~Beisert and M.~Staudacher, ``{The ${\mathcal N}=4$ SYM integrable super spin
  chain},'' {\em Nucl. Phys.} {\bf B670} (2003) 439--463,
\href{http://arXiv.org/abs/hep-th/0307042}{{\tt hep-th/0307042}}.

\bibitem{Kazakov:2004qf}
V.~A. Kazakov, A.~Marshakov, J.~A. Minahan, and K.~Zarembo, ``{Classical /
  quantum integrability in AdS/CFT},'' {\em JHEP} {\bf 05} (2004) 024,
\href{http://arXiv.org/abs/hep-th/0402207}{{\tt hep-th/0402207}}.

\bibitem{Beisert:2004hm}
N.~Beisert, V.~Dippel, and M.~Staudacher, ``{A novel long range spin chain and
  planar ${\mathcal N}=4$ super Yang- Mills},'' {\em JHEP} {\bf 07} (2004) 075,
\href{http://arXiv.org/abs/hep-th/0405001}{{\tt hep-th/0405001}}.

\bibitem{Staudacher:2004tk}
M.~Staudacher, ``{The factorized S-matrix of CFT/AdS},'' {\em JHEP} {\bf 05}
  (2005) 054,
\href{http://arXiv.org/abs/hep-th/0412188}{{\tt hep-th/0412188}}.

\bibitem{Arutyunov:2004vx}
G.~Arutyunov, S.~Frolov, and M.~Staudacher, ``{Bethe ansatz for quantum
  strings},'' {\em JHEP} {\bf 10} (2004) 016,
\href{http://arXiv.org/abs/hep-th/0406256}{{\tt hep-th/0406256}}.

\bibitem{Beisert:2005fw}
N.~Beisert and M.~Staudacher, ``{Long-range $\mathfrak{psu}(2,2|4)$ Bethe
  ansaetze for gauge theory and strings},'' {\em Nucl. Phys.} {\bf B727} (2005)
  1--62,
\href{http://arXiv.org/abs/hep-th/0504190}{{\tt hep-th/0504190}}.

\bibitem{Beisert:2005tm}
N.~Beisert, ``{The $\mathfrak{su}(2|2)$ dynamic S-matrix},''
\href{http://arXiv.org/abs/hep-th/0511082}{{\tt hep-th/0511082}}.

\bibitem{Janik:2006dc}
R.~A. Janik, ``{The $AdS_{5}\times S^{5}$ superstring worldsheet S-matrix and
  crossing symmetry},'' {\em Phys. Rev.} {\bf D73} (2006) 086006,
\href{http://arXiv.org/abs/hep-th/0603038}{{\tt hep-th/0603038}}.

\bibitem{Eden:2006rx}
B.~Eden and M.~Staudacher, ``{Integrability and transcendentality},'' {\em J.
  Stat. Mech.} {\bf 0611} (2006) P014,
\href{http://arXiv.org/abs/hep-th/0603157}{{\tt hep-th/0603157}}.

\bibitem{Arutyunov:2006iu}
G.~Arutyunov and S.~Frolov, ``{On $AdS_{5}\times S^{5}$ string S-matrix},''
  {\em Phys. Lett.} {\bf B639} (2006) 378--382,
\href{http://arXiv.org/abs/hep-th/0604043}{{\tt hep-th/0604043}}.

\bibitem{Beisert:2006ib}
N.~Beisert, R.~Hernandez, and E.~Lopez, ``{A crossing-symmetric phase for
  $AdS_{5}\times S^{5}$ strings},'' {\em JHEP} {\bf 11} (2006) 070,
\href{http://arXiv.org/abs/hep-th/0609044}{{\tt hep-th/0609044}}.

\bibitem{Beisert:2006ez}
N.~Beisert, B.~Eden, and M.~Staudacher, ``{Transcendentality and crossing},''
  {\em J. Stat. Mech.} {\bf 0701} (2007) P021,
\href{http://arXiv.org/abs/hep-th/0610251}{{\tt hep-th/0610251}}.

\bibitem{Arutyunov:2006yd}
G.~Arutyunov, S.~Frolov, and M.~Zamaklar, ``{The Zamolodchikov-Faddeev algebra
  for $AdS_{5}\times S^{5}$ superstring},'' {\em JHEP} {\bf 04} (2007) 002,
\href{http://arXiv.org/abs/hep-th/0612229}{{\tt hep-th/0612229}}.

\bibitem{Dorey:2007xn}
N.~Dorey, D.~M. Hofman, and J.~Maldacena, ``{On the singularities of the magnon
  S-matrix},'' {\em Phys. Rev.} {\bf D76} (2007) 025011,
\href{http://arXiv.org/abs/hep-th/0703104}{{\tt hep-th/0703104}}.

\bibitem{Berenstein:2002jq}
D.~Berenstein, J.~M. Maldacena, and H.~Nastase, ``{Strings in flat space and pp
  waves from ${\mathcal N}=4$ super Yang Mills},'' {\em JHEP} {\bf 04} (2002)
  013,
\href{http://arXiv.org/abs/hep-th/0202021}{{\tt hep-th/0202021}}.

\bibitem{Rej:2007vm}
A.~Rej, M.~Staudacher, and S.~Zieme, ``{Nesting and dressing},''
\href{http://arXiv.org/abs/hep-th/0702151}{{\tt hep-th/0702151}}.

\bibitem{Janik:2008hs}
R.~A. Janik and T.~Lukowski, ``{From nesting to dressing},'' {\em Phys. Rev.}
  {\bf D78} (2008) 066018,
\href{http://arXiv.org/abs/0804.4295}{{\tt 0804.4295}}.

\bibitem{Gubser:2002tv}
S.~S. Gubser, I.~R. Klebanov, and A.~M. Polyakov, ``{A semi-classical limit of
  the gauge/string correspondence},'' {\em Nucl. Phys.} {\bf B636} (2002)
  99--114,
\href{http://arXiv.org/abs/hep-th/0204051}{{\tt hep-th/0204051}}.

\bibitem{FT}
S.~Frolov and A.~A. Tseytlin, ``{Multi-spin string solutions in $AdS_{5}\times
  S^{5}$},'' {\em Nucl. Phys.} {\bf B668} (2003) 77--110,
\href{http://arXiv.org/abs/hep-th/0304255}{{\tt hep-th/0304255}}.
~{\small $\bullet$}~
S.~Frolov and A.~A. Tseytlin, ``{Quantizing three-spin string solution in
  $AdS_{5}\times S^{5}$},'' {\em JHEP} {\bf 07} (2003) 016,
\href{http://arXiv.org/abs/hep-th/0306130}{{\tt hep-th/0306130}}.

\bibitem{Hofman:2006xt}
D.~M. Hofman and J.~M. Maldacena, ``{Giant magnons},'' {\em J. Phys.} {\bf A39}
  (2006) 13095--13118,
\href{http://arXiv.org/abs/hep-th/0604135}{{\tt hep-th/0604135}}.

\bibitem{Ishizeki:2007we}
R.~Ishizeki and M.~Kruczenski, ``{Single spike solutions for strings on $S^{2}$
  and $S^{3}$},''
\href{http://arXiv.org/abs/arXiv:0705.2429 [hep-th]}{{\tt arXiv:0705.2429
  [hep-th]}}.

\bibitem{Mosaffa:2007ty}
A.~E. Mosaffa and B.~Safarzadeh, ``{Dual Spikes: New Spiky String Solutions},''
\href{http://arXiv.org/abs/arXiv:0705.3131 [hep-th]}{{\tt arXiv:0705.3131
  [hep-th]}}.

\bibitem{Roiban:2006jt}
R.~Roiban, A.~Tirziu, and A.~A. Tseytlin, ``{Slow-string limit and
  'antiferromagnetic' state in AdS/CFT},'' {\em Phys. Rev.} {\bf D73} (2006)
  066003,
\href{http://arXiv.org/abs/hep-th/0601074}{{\tt hep-th/0601074}}.

\bibitem{Zarembo:2005ur}
K.~Zarembo, ``{Antiferromagnetic operators in $\mathcal N = 4$ 
supersymmetric Yang-Mills theory},'' {\em Phys. Lett.} {\bf B634} (2006)
  552-556,
\href{http://arXiv.org/abs/hep-th/0512079}{{\tt hep-th/0512079}}.

\bibitem{Rej:2005qt}
A.~Rej, D.~Serban, and M.~Staudacher, ``{Planar ${\mathcal N}=4$ gauge theory
  and the Hubbard model},'' {\em JHEP} {\bf 03} (2006) 018,
\href{http://arXiv.org/abs/hep-th/0512077}{{\tt hep-th/0512077}}.

\bibitem{Hayashi:2007bq}
H.~Hayashi, K.~Okamura, R.~Suzuki, and B.~Vicedo, ``{Large Winding Sector of
  AdS/CFT},'' {\em JHEP} {\bf 11} (2007) 033,
\href{http://arXiv.org/abs/arXiv:0709.4033 [hep-th]}{{\tt arXiv:0709.4033
  [hep-th]}}.

\bibitem{Ishizeki:2007kh}
R.~Ishizeki, M.~Kruczenski, M.~Spradlin, and A.~Volovich, ``{Scattering of
  single spikes},''
\href{http://arXiv.org/abs/arXiv:0710.2300 [hep-th]}{{\tt arXiv:0710.2300
  [hep-th]}}.

\bibitem{Abbott:2008yp}
M.~C. Abbott and I.~V. Aniceto, ``{Vibrating giant spikes and the large-winding
  sector},'' {\em JHEP} {\bf 06} (2008) 088,
\href{http://arXiv.org/abs/0803.4222}{{\tt 0803.4222}}.

\bibitem{Ahn:2008sk}
C.~Ahn and P.~Bozhilov, ``{Finite-size Effects for Single Spike},'' {\em JHEP}
  {\bf 07} (2008) 105,
\href{http://arXiv.org/abs/0806.1085}{{\tt 0806.1085}}.

\bibitem{Minahan:2006bd}
J.~A. Minahan, A.~Tirziu, and A.~A. Tseytlin, ``{Infinite spin limit of
  semiclassical string states},'' {\em JHEP} {\bf 08} (2006) 049,
\href{http://arXiv.org/abs/hep-th/0606145}{{\tt hep-th/0606145}}.

\bibitem{Okamura:2006zv}
K.~Okamura and R.~Suzuki, ``{A perspective on classical strings from complex
  sine-Gordon solitons},'' {\em Phys. Rev.} {\bf D75} (2007) 046001,
\href{http://arXiv.org/abs/hep-th/0609026}{{\tt hep-th/0609026}}.

\bibitem{Beisert:2005di}
N.~Beisert, V.~A. Kazakov, K.~Sakai, and K.~Zarembo, ``{Complete spectrum of
  long operators in $\mathcal N = 4$ SYM at one loop},'' {\em JHEP} {\bf 07}
  (2005) 030,
\href{http://arXiv.org/abs/hep-th/0503200}{{\tt hep-th/0503200}}.

\bibitem{Lin:2004nb}
H.~Lin, O.~Lunin, and J.~M. Maldacena, ``{Bubbling AdS space and 1/2 BPS
  geometries},'' {\em JHEP} {\bf 10} (2004) 025,
\href{http://arXiv.org/abs/hep-th/0409174}{{\tt hep-th/0409174}}.

\bibitem{dressing}
M.~Spradlin and A.~Volovich, ``{Dressing the giant magnon},'' {\em JHEP} {\bf
  10} (2006) 012,
\href{http://arXiv.org/abs/hep-th/0607009}{{\tt hep-th/0607009}}.
~{\small $\bullet$}~
C.~Kalousios, M.~Spradlin, and A.~Volovich, ``{Dressing the giant magnon.
  II},'' {\em JHEP} {\bf 03} (2007) 020,
\href{http://arXiv.org/abs/hep-th/0611033}{{\tt hep-th/0611033}}.

\bibitem{scatt:boundstate}
H.-Y. Chen, N.~Dorey, and K.~Okamura, ``{On the scattering of magnon
  boundstates},'' {\em JHEP} {\bf 11} (2006) 035,
\href{http://arXiv.org/abs/hep-th/0608047}{{\tt hep-th/0608047}}.
~{\small $\bullet$}~
R.~Roiban, ``{Magnon bound-state scattering in gauge and string theory},'' {\em
  JHEP} {\bf 04} (2007) 048,
\href{http://arXiv.org/abs/hep-th/0608049}{{\tt hep-th/0608049}}.

\bibitem{boundstate}
N.~Dorey, ``{Magnon bound states and the AdS/CFT correspondence},'' {\em J.
  Phys.} {\bf A39} (2006) 13119--13128,
\href{http://arXiv.org/abs/hep-th/0604175}{{\tt hep-th/0604175}}.
~{\small $\bullet$}~
H.-Y. Chen, N.~Dorey, and K.~Okamura, ``{Dyonic giant magnons},'' {\em JHEP}
  {\bf 09} (2006) 024,
\href{http://arXiv.org/abs/hep-th/0605155}{{\tt hep-th/0605155}}.
~{\small $\bullet$}~
H.-Y. Chen, N.~Dorey, and K.~Okamura, ``{The asymptotic spectrum of the
  ${\mathcal N}=4$ super Yang-Mills spin chain},'' {\em JHEP} {\bf 03} (2007)
  005,
\href{http://arXiv.org/abs/hep-th/0610295}{{\tt hep-th/0610295}}.
~{\small $\bullet$}~
G.~Arutyunov, S.~Frolov, and M.~Zamaklar, ``{Finite-size effects from giant
  magnons},'' {\em Nucl. Phys.} {\bf B778} (2007) 1--35,
\href{http://arXiv.org/abs/hep-th/0606126}{{\tt hep-th/0606126}}.

\bibitem{NLIE}
G.~Feverati, D.~Fioravanti, P.~Grinza, and M.~Rossi, ``{On the finite size
  corrections of anti-ferromagnetic anomalous dimensions in $\cN = 4$ SYM},''
  {\em JHEP} {\bf 05} (2006) 068,
\href{http://arXiv.org/abs/hep-th/0602189}{{\tt hep-th/0602189}}.
~{\small $\bullet$}~
G.~Feverati, D.~Fioravanti, P.~Grinza, and M.~Rossi, ``{Hubbard's adventures in
  $\cN = 4$ SYM-land? Some non- perturbative considerations on finite length
  operators},'' {\em J. Stat. Mech.} {\bf 0702} (2007) P001,
\href{http://arXiv.org/abs/hep-th/0611186}{{\tt hep-th/0611186}}.
~{\small $\bullet$}~
D.~Fioravanti and M.~Rossi, ``{On the commuting charges for the highest
  dimension $SU(2)$ operator in planar $\cN=4$ SYM},'' {\em JHEP} {\bf 08}
  (2007) 089,
\href{http://arXiv.org/abs/0706.3936}{{\tt 0706.3936}}.

\bibitem{Aharony:2008ug}
O.~Aharony, O.~Bergman, D.~L. Jafferis, and J.~Maldacena, ``{$\cN = 6$
  superconformal Chern-Simons-matter theories, M2-branes and their gravity
  duals},'' {\em JHEP} {\bf 10} (2008) 091,
\href{http://arXiv.org/abs/0806.1218}{{\tt 0806.1218}}.

\bibitem{AdS4CFT3}
J.~A. Minahan and K.~Zarembo, ``{The Bethe ansatz for superconformal
  Chern-Simons},'' {\em JHEP} {\bf 09} (2008) 040,
\href{http://arXiv.org/abs/0806.3951}{{\tt 0806.3951}}.
~{\small $\bullet$}~
N.~Gromov and P.~Vieira, ``{The all loop AdS${}_{4}$/CFT${}_{3}$ Bethe
  ansatz},'' {\em JHEP} {\bf 01} (2009) 016,
\href{http://arXiv.org/abs/0807.0777}{{\tt 0807.0777}}.
~{\small $\bullet$}~
N.~Gromov and P.~Vieira, ``{The AdS${}_{4}$/CFT${}_{3}$ algebraic curve},''
  {\em JHEP} {\bf 02} (2009) 040,
\href{http://arXiv.org/abs/0807.0437}{{\tt 0807.0437}}.
~{\small $\bullet$}~
D.~Astolfi, V.~G.~M. Puletti, G.~Grignani, T.~Harmark, and M.~Orselli,
  ``{Finite-size corrections in the $SU(2)\times SU(2)$ sector of type IIA
  string theory on ${\rm AdS}_{4}\times\mathbb{CP}^{3}$},'' {\em Nucl. Phys.}
  {\bf B810} (2009) 150--173,
\href{http://arXiv.org/abs/0807.1527}{{\tt 0807.1527}}.
~{\small $\bullet$}~
C.~Ahn and R.~I. Nepomechie, ``{$\cN = 6$ super Chern-Simons theory S-matrix
  and all-loop Bethe ansatz equations},'' {\em JHEP} {\bf 09} (2008) 010,
\href{http://arXiv.org/abs/0807.1924}{{\tt 0807.1924}}.
~{\small $\bullet$}~
D.~Bak and S.-J. Rey, ``{Integrable Spin Chain in Superconformal Chern-Simons
  Theory},'' {\em JHEP} {\bf 10} (2008) 053,
\href{http://arXiv.org/abs/0807.2063}{{\tt 0807.2063}}.
~{\small $\bullet$}~
T.~McLoughlin and R.~Roiban, ``{Spinning strings at one-loop in ${\rm
  AdS}_{4}\times\mathbb{P}^{3}$},'' {\em JHEP} {\bf 12} (2008) 101,
\href{http://arXiv.org/abs/0807.3965}{{\tt 0807.3965}}.
~{\small $\bullet$}~
T.~McLoughlin, R.~Roiban, and A.~A. Tseytlin, ``{Quantum spinning strings in
  ${\rm AdS}_{4}\times\mathbb{CP}^{3}$: testing the Bethe Ansatz proposal},''
  {\em JHEP} {\bf 11} (2008) 069,
\href{http://arXiv.org/abs/0809.4038}{{\tt 0809.4038}}.
~{\small $\bullet$}~
L.~F. Alday, G.~Arutyunov, and D.~Bykov, ``{Semiclassical Quantization of
  Spinning Strings in ${\rm AdS}_{4}\times\mathbb{CP}^{3}$},'' {\em JHEP} {\bf
  11} (2008) 089,
\href{http://arXiv.org/abs/0807.4400}{{\tt 0807.4400}}.
~{\small $\bullet$}~
C.~Krishnan, ``{AdS$_{4}$/CFT$_{3}$ at One Loop},'' {\em JHEP} {\bf 09} (2008)
  092,
\href{http://arXiv.org/abs/0807.4561}{{\tt 0807.4561}}.
~{\small $\bullet$}~
N.~Gromov and V.~Mikhaylov, ``{Comment on the Scaling Function in ${\rm
  AdS}_{4}\times\mathbb{CP}^{3}$},'' {\em JHEP} {\bf 04} (2009) 083,
\href{http://arXiv.org/abs/0807.4897}{{\tt 0807.4897}}.

\bibitem{AdS4CFT3:giants}
M.~C. Abbott and I.~Aniceto, ``{Giant Magnons in ${\rm
  AdS}_{4}\times\mathbb{CP}^{3}$: Embeddings, Charges and a Hamiltonian},''
  {\em JHEP} {\bf 04} (2009) 136,
\href{http://arXiv.org/abs/0811.2423}{{\tt 0811.2423}}.
~{\small $\bullet$}~
I.~Shenderovich, ``{Giant magnons in ${\rm AdS}_{4}/{\rm CFT}_{3}$: dispersion,
  quantization and finite-size corrections},''
\href{http://arXiv.org/abs/0807.2861}{{\tt 0807.2861}}.
~{\small $\bullet$}~
C.~Ahn, P.~Bozhilov, and R.~C. Rashkov, ``{Neumann-Rosochatius integrable
  system for strings on ${\rm AdS}_{4}\times\mathbb{CP}^{3}$},'' {\em JHEP}
  {\bf 09} (2008) 017,
\href{http://arXiv.org/abs/0807.3134}{{\tt 0807.3134}}.
~{\small $\bullet$}~
C.~Ahn and P.~Bozhilov, ``{Finite-size Effect of the Dyonic Giant Magnons in
  $\cN=6$ super Chern-Simons Theory},'' {\em Phys. Rev.} {\bf D79} (2009)
  046008,
\href{http://arXiv.org/abs/0810.2079}{{\tt 0810.2079}}.
~{\small $\bullet$}~
M.~C. Abbott, I.~Aniceto, and O.~O. Sax, ``{Dyonic Giant Magnons in
  $\mathbb{CP}^{3}$: Strings and Curves at Finite J},'' {\em Phys. Rev.} {\bf
  D80} (2009) 026005,
\href{http://arXiv.org/abs/0903.3365}{{\tt 0903.3365}}.
~{\small $\bullet$}~
D.~Bombardelli and D.~Fioravanti, ``{Finite-Size Corrections of the
  $\mathbb{CP}^{3}$ Giant Magnons: the L\'{u}scher terms},'' {\em JHEP} {\bf
  07} (2009) 034,
\href{http://arXiv.org/abs/0810.0704}{{\tt 0810.0704}}.
~{\small $\bullet$}~
T.~Lukowski and O.~O. Sax, ``{Finite size giant magnons in the $SU(2)\times
  SU(2)$ sector of ${\rm AdS}_{4}\times\mathbb{CP}^{3}$},'' {\em JHEP} {\bf 12}
  (2008) 073,
\href{http://arXiv.org/abs/0810.1246}{{\tt 0810.1246}}.
~{\small $\bullet$}~
G.~Grignani, T.~Harmark, and M.~Orselli, ``{The $SU(2)\times SU(2)$ sector in
  the string dual of $\cN=6$ superconformal Chern-Simons theory},'' {\em Nucl.
  Phys.} {\bf B810} (2009) 115--134,
\href{http://arXiv.org/abs/0806.4959}{{\tt 0806.4959}}.
~{\small $\bullet$}~
G.~Grignani, T.~Harmark, M.~Orselli, and G.~W. Semenoff, ``{Finite size Giant
  Magnons in the string dual of $\cN=6$ superconformal Chern-Simons theory},''
  {\em JHEP} {\bf 12} (2008) 008,
\href{http://arXiv.org/abs/0807.0205}{{\tt 0807.0205}}.
~{\small $\bullet$}~
B.-H. Lee, K.~L. Panigrahi, and C.~Park, ``{Spiky Strings on $AdS_{4}\times
  {\mathbb{CP}}^{3}$},'' {\em JHEP} {\bf 11} (2008) 066,
\href{http://arXiv.org/abs/0807.2559}{{\tt 0807.2559}}.
~{\small $\bullet$}~
T.~J. Hollowood and J.~L. Miramontes, ``{Magnons, their Solitonic Avatars and
  the Pohlmeyer Reduction},'' {\em JHEP} {\bf 04} (2009) 060,
\href{http://arXiv.org/abs/0902.2405}{{\tt 0902.2405}}.
~{\small $\bullet$}~
R.~Suzuki, ``{Giant Magnons on $\mathbb{CP}^{3}$ by Dressing Method},'' {\em
  JHEP} {\bf 05} (2009) 079,
\href{http://arXiv.org/abs/0902.3368}{{\tt 0902.3368}}.
~{\small $\bullet$}~
C.~Kalousios, M.~Spradlin, and A.~Volovich, ``{Dyonic Giant Magnons on
  $\mathbb{CP}^{3}$},'' {\em JHEP} {\bf 07} (2009) 006,
\href{http://arXiv.org/abs/0902.3179}{{\tt 0902.3179}}.

\end{thebibliography}

\end{document}